\documentclass[preprint, aps,amsmath,amsfonts,amssymb]{revtex4-1}

\usepackage{epsfig}
\usepackage{graphicx}
\usepackage{hyperref}
\usepackage{epstopdf}
 \usepackage[utf8]{inputenc}
\usepackage[]{natbib}

\newcommand{\be}{\begin{equation}}
\newcommand{\ee}{\end{equation}}
\newcommand{\bea}{\begin{eqnarray}}
\newcommand{\eea}{\end{eqnarray}}
\newcommand{\ba}{\begin{array}}
\newcommand{\ea}{\end{array}}
\newcommand{\la}{\begin{langle}}
\newcommand{\ra}{\end{rangle}}

\begin{document}

\title{Momentum transfer dependence of generalized parton distributions }
\author{Neetika Sharma }

\affiliation{
Indian Institute of Science Education and Research Mohali,\\
S.A.S. Nagar, Mohali-140306, Punjab, India.
}

\begin{abstract}
We revisit the model for parametrization of momentum  dependence of nucleon generalized parton distributions  in the light of recent MRST measurements of parton distribution functions \cite{Martin:2009iq}.  Our parametrization  method with minimum set of free parameters give a sufficiently good description of data for Dirac and Pauli electromagnetic form factors of  proton and neutron at small and intermediate values of  momentum transfer. We  also calculate the GPDs for up and down quark by decomposing the electromagnetic form factors for  nucleon using the charge and isospin symmetry and also study the evolution of GPDs to a higher scale. We further investigate  the transverse charge densities for both the unpolarized and transversely polarized nucleon and compare our  results with  the Kelly's distribution.  
\end{abstract}
\maketitle


\section{Introduction} 

The study of fundamental quantities like Generalized partion distributions (GPDs) and electromagnetic form  factors (EFFs)  have been of tremendous interest in the recent past. GPDs are parametrized in terms of longitudinal momentum $x$ and invariant momentum transfer squared $t=-q^2$,  thus giving  vital information about the  3-D structure of nucleon. The EFFs measure the probability for a nucleon to absorb a virtual photon of momentum  $q^2$. Ji sum rule  state that the GPD for unpolarized nucleon reduce to EFFs when integrated over $x$ \cite{jisum} :
\bea \label{sumrules}
F_1^N(q^2) &=& \int_{0}^1 dx\;    H^N(x, q^2) \,, \\
F_2^N(q^2) &=& \int_{0}^1 dx\;   E^N(x, q^2) \,, \eea
where $H(x,q^2)$ and  $E(x,q^2)$ (with $N={p,n}$) are, respectively, the spin independent and spin dependent GPDs  and the functions $F_1$ and $F_2$ are Dirac and Pauli form factors for nucleon.

GPDs are measured in the processes, like deeply virtual Compton scattering  $(\gamma^* p \to \gamma p)$ and diffractive vector meson production $(\gamma^* p \to V p)$.  Various experiments at Jefferson Lab (JLAB) \cite{jlab,jlab1}, and HERA \cite{hermes} have extracted information about GPDs and new data  is expected in near future from the experiments at JLAB12, COMPASS, and electron-ion collider  \cite{compass}. On the other hand, GPDs are calculated in the Euclidean lattice QCD using the first principle approach \cite{lattice1,*lattice2}. This is a method based upon numerical simulations,  however, the applicability is limited by enormous computational complexities, dependence upon the free parameters, and the fact that dynamical observables in Minkowski spacetime can not be directly observed in the Euclidean lattice computations.  Recently, there has been a lot of activity in  calculating GPDs in the  anti-de Sitter AdS/QCD holography based models as well \cite{hard,*soft}. In this approach, information on  GPDs is obtained via matching the matrix elements of  electromagnetic current  with QCD sum rules in the hard and soft-wall models \cite{abidin,*stan2012}.

There are several other phenomenological models in literature to obtain GPDs in a functional form that represent the data over a wide range of momentum transferred \cite{kelly,gaussian,poly1,*poly2,regge,etc}. These approaches are based on parametrizations of the quark wave function or directly the EFFs, using constraint  imposed by sum rules in Eq.(\ref{sumrules}).  It is also interesting to formulate simple parametrization in term of $x$  and $q^2$-dependences.  The $x$-dependence were measured accurately in terms of the quark distribution functions  and $q^2$-dependence can be added using different functional forms  from the phenomenological description of nucleon EFFs.

In this work, we will follow  parametrization approach with a functional form of  quark distribution functions based on the latest global analysis by ``MSTW2009''  \cite{Martin:2009iq} combined with  the guassian ansatz  to incorporate $q^2$-dependence \cite{oleg}. Though this approach is model dependent, but on the contrary  one can reach the much higher values of momentum transferred, even larger than measured in hard exclusive scattering experiments.

The  purpose of  present communication  is to give a simple parametrization of spin flip and non flip nucleon GPDs ($H$ and $E$) with a minimum number of free parameters.  This parametrization can serve as a baseline for comparison with other GPD fits, for the interpolation of data on FFs and calculations of transverse charge and magnetization densities. The best fit set of parameters are obtained after a critical examination of the available measurements of EFFs. Further, we extracted the individual contributions of up and down quarks to the Dirac and Pauli form factors and presented a comparison with the experimental data.   In order to check the consistency of our method, we  also presented results for the perturbative evolution of GPDs according to the Dokshitzer-Gribov-Lipatov-Altarelli-Parisi-like (DGLAP) equations for the up and down quarks in momentum space.

It is also interesting to investigate the transverse charge and magnetization densities  as they provide insight into the structure of the nucleon in the valence region. The charge and magnetization densities in the transverse plane are defined as the Fourier transform of the EFFs.  We therefore, investigated the  charge and magnetization densities for both unpolarized and transversely polarized nucleon  in the parametrization approach and present a comparison with the Kelly's method.


\section{ Momentum  dependence of GPDs in Parametrization approach}
\label{para}

A large part of  present day high-energy scattering experiments, such as, proton-proton  (proton-antiproton) collisions and deep-inelastic scattering, proceed via the  knowledge of  partonic constituents of the hadron called parton distribution functions (PDFs).  Adding the $q^2$-dependence in PDFs gives the 3-D structure of hadrons in terms of the GPDs. The Fourier transform of GPDs provide information about the distribution of the  charge and magnetization densities of quarks in impact parameter space. As we cannot obtain $q^2$-dependence from the first principles, it would be interesting to formulate a simple parametrization that accurately represent the data on FFs over a wide range of momentum transferred.

There are various phenomenological parametrizations for the extraction of GPDs  in the literature \cite{kelly}.  The simplest model for parametrization of  proton GPDs assumes a gaussian form of wavefunction  with an interplay between $x$ and $q^2$-dependences \cite{gaussian}. Further, a Regge parametrization for GPDs $H(x,q) =q(x)  \exp[- \alpha\, q^2] $ is used  at small momentum transfer \cite{Goeke},  and  a modified version  of profile function $q(x) \exp[- \alpha' (1-x) \, q^2]$ is used to extend the analysis to large momentum transfer region  \cite{gaussian}. Some of the  other well known procedures are based on  polynomial or logarithmic form  of  profile  function in accordance  with theoretical and phenomenological constraints  \cite{poly1,poly2}. These approaches give a satisfactory description of the basic features of proton and neutron EFFs data.

In a recent parametrization method (PM) \cite{oleg},  $q^2$-dependence  is incorporated by combining the  gaussian ansatz with the information on PDFs obtained from  global fit to the experimental data, however, these calculation are being performed using the older set of PDFs by MRST2002  fit  \cite{Martin:2002dr}.  Recently the  calculations of PDFs have been revisited in the light of  considerable improvements in  experimental precision, wider kinematic range, as well as the availability of new data. In view of these developments, it is interesting to revisit the calculations on parametrization of GPDs using the new set  of PDFs  proposed  in the ``MSTW2009'' global fit \cite{Martin:2009iq}.

GPD for Dirac form factor for the case of proton  \be {H}^p (x,q^2)= \sum_q e_q \, q(x) \, \exp \left[{-a_q^p {(1-x)^2  \over x^{m_p} } q^2} \right] \,, \label{hp} \ee where $e_q={2 / 3 }$ for up and $-{1/ 3}$ for down quark and  $a^p_u$, $a^p_d$ and $m_p$ are the free parameters to be fitted from the low $q^2$ experimental data on the proton form factors.  The quark distribution functions $q(x)$ measured at the next-to-next-to-leading-order in strong coupling parameter by ``MSTW2009''  (at initial scale $\mu_o^2$=1 GeV$^2$) are expressed as  \cite{Martin:2009iq}
\bea 
x\,u(x) &=& 0.22x^{0.28} (1-x)^{3.36}  (1 + 4.43 \sqrt{x}  + 38.6\, x )\,,  \\
x\,d(x) &=& 17.94 x^{1.08} (1-x)^{6.15} (1- 3.64 \sqrt{x} +  5.26 \,x )\,.  
\eea

We parametrize the GPD $E^p(x,q^2)$ for the proton using the widely used representation  \cite{gaussian}: \be {E}^p (x,q^2)= \sum_q e_q  {\cal E}_q (x) \, \exp \left[{-a^p_q { (1-x)^2  \over  x^{m_p} } q^2 } \right] \,, \ee with \bea  {\cal  E}_u(x) = {\kappa_u \over N_u} (1-x)^{\kappa_1} \, u(x)\,,\\ {\cal E}_d(x) = {\kappa_d \over N_d} (1-x)^{\kappa_2} \, d(x)\,, \eea
where the normalization of up and down quark GPDs to their corresponding anomalous magnetic moments $\kappa_u = 1.673$, $\kappa_d= -2.033$ lead to the $k_1=1.53$, $k_2=0.31$, $N_u =1.52$ and $N_d = 0.95$. We follow the similar kind of parametrization for neutron GPDs while invoking the charge symmetry  and changing the index  $p$ to $n$ for free parameters. Though we have kept the minimal structure of parametrization same as previous work  \cite{oleg}, we have increased the number of free parameters in order to give a better fit to the data over a wider range.

The experimental results are usually expressed in terms of the Sachs form factors $G^{p,n}_M(q^2)$ and  $G^{p,n}_E(q^2)$.  Further, the Dirac and Pauli form factor are  related to Sachs form factor by relation: \bea  G_M^{p,n} &=& F_1^{p,n} +F_2^{p,n} \,, \\G_E^{p,n} &=& F_1^{p,n} -{q^2 \over 4 M_N^2}F_2^{p,n} \,, \eea where $M_N$ is the mass of proton and neutron.  Several experiments measure the ratio of electric and magnetic form factors, which is commonly defined as $ R^i(q^2) = \mu_i G_E^i (q^2) / G^i_M (q^2)\,$ (with ${i=p,n}$). The magnetic moments $\mu_p$  and $\mu_n$ normalize the ratio to unity at zero momentum transferred. For convenience the electric and magnetic form factors are often divided by the conventional dipole form $ G_D(t) = { 1 / ( 1 +q^2 /\Lambda^2 )^2} \,,$ with $\Lambda^2=0.71 $ {GeV}$^2$.

We now briefly discuss the various input parameters needed in the numerical calculations. Our approach involves six free parameters, $a_u^p$ , $a_d^p$, and $m^p$ for proton and $a_u^n$ , $a_d^n$ and $m_n$ for neutron. In order to  find the best fit set, we have used  the latest data available on the proton and neutron form factors.  We have used the data $G_E(q^2)/G_D(q^2) $ \cite{global},  $ {G_M (q^2) /(\mu_p G_D (q^2))}$ \cite{global}, and  $R_p$ \cite{milbrath,jones,gayou,gayou2,punjabi,hu,crawford,global,maclachlan, paolone,puckett,puckett2} for the case proton. The best fit set of parameters are $a^p_u= 1.21$, $a^p_d= 1.41$, and $m_p=0.35$. For the case of neutron, we have used the $G_E (q^2)$ \cite{Glazier:2004ny,plaster,riordan}, and  $R_n$ \cite{riordan,geis,schlimme} and the best fit set of parameter are  $a^n_u= 1.49$, $a^n_d=1.66$, and $m_n=0.26$. We have also added the systematic and  statistical uncertainties in  quadrature for the uniform treatment of data wherever they are available.

\begin{figure}[htbt]
\begin{center}
\includegraphics[width=15cm,height=11cm,clip]{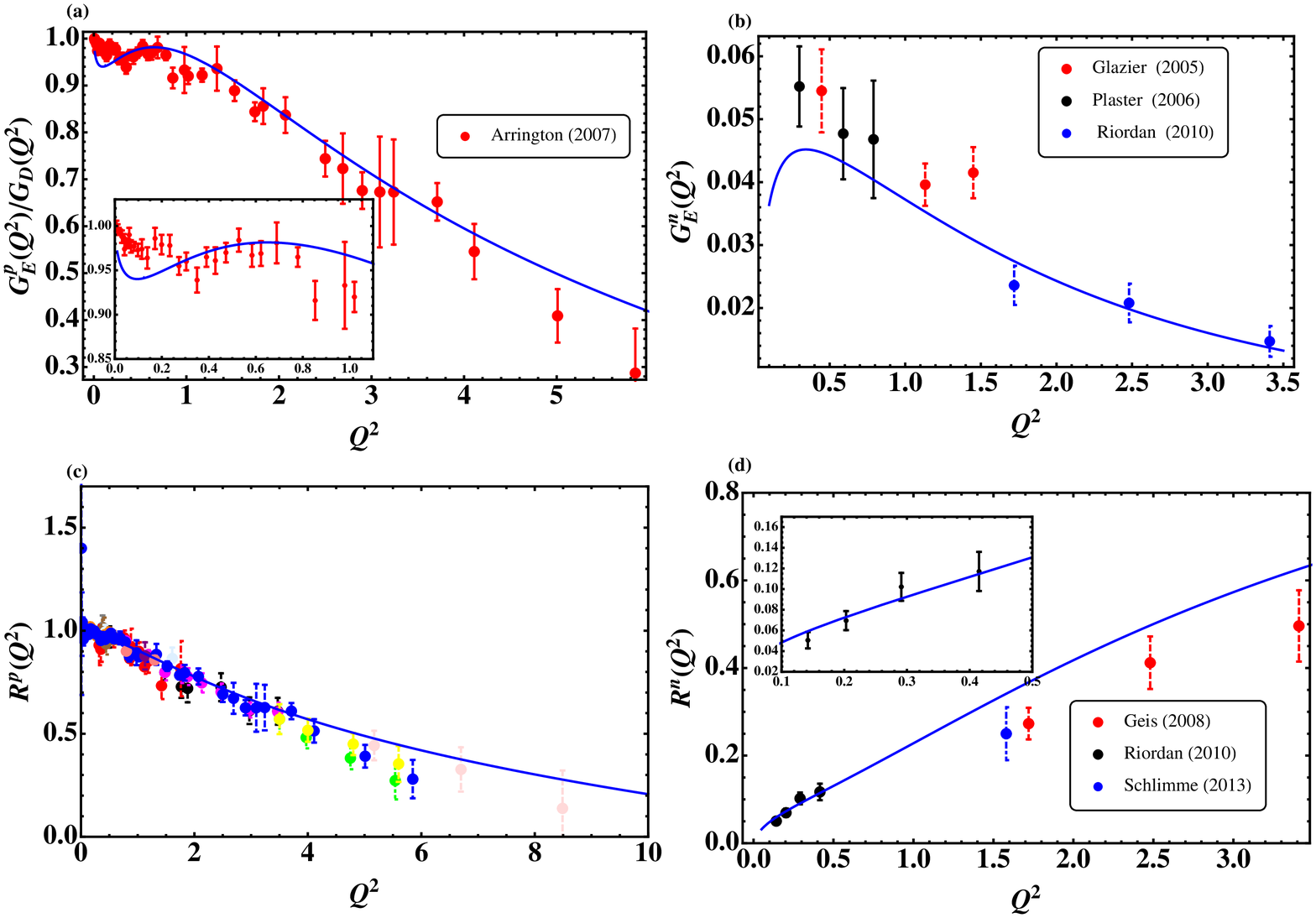}
\caption{\label{gep1} (a) Plots for  proton  electric $G^p_E$ form factors  divided by the dipole $G_D = 1/(1+ q^2/0.71 ~{\rm GeV}^2)^2 $ plotted vs  $q^2$ (b) the neutron electric form factors $G^n_E$ plotted vs  $q^2$, (c-d) represent the plots for the ratio of  electric to the magnetic form factors vs $q^2$ for proton and neutron, respectively. The experimental data points for $R_p$ are obtained  from Ref.  \cite{milbrath,jones,gayou,gayou2,punjabi,hu,crawford,global,maclachlan, paolone,puckett,puckett2}. Solid blue line  correspond to the results obtained in the parametrization method. }
\end{center}
\end{figure}

In Fig. \ref{gep1}(a)-(b),  we presented the comparison of Sach form factors $G^p_E(q^2)$ and $G_E^n(q^2)$  with $q^2$ for proton and neutron in the parametrization method.  The data for the proton form factor is  divided by the dipole form  $G_D (q^2)$  to make a direct comparison with the experimental data \cite{global}.  Fig. \ref{gep1}(c)-(d), predict the behavior of ratio of electric to magnetic form factors  $R_p$  and  $R_n$   with $q^2$ for proton and neutron, respectively. In all the cases, we have presented a comparison with the experimental data  \cite{riordan,plaster,geis, milbrath,jones,gayou,gayou2,punjabi, hu,crawford,global,maclachlan,paolone,puckett,puckett2, Glazier:2004ny, plaster,riordan, geis, schlimme}. It can be easily seen that a basic variant of model with minimum number of free parameters lead to a very  good description of the $q^2$-dependence of the data at both low and intermediate $q^2$ values. 

\begin{figure}[htbt]
\begin{center}
\includegraphics[width=15cm,height=11cm,clip]{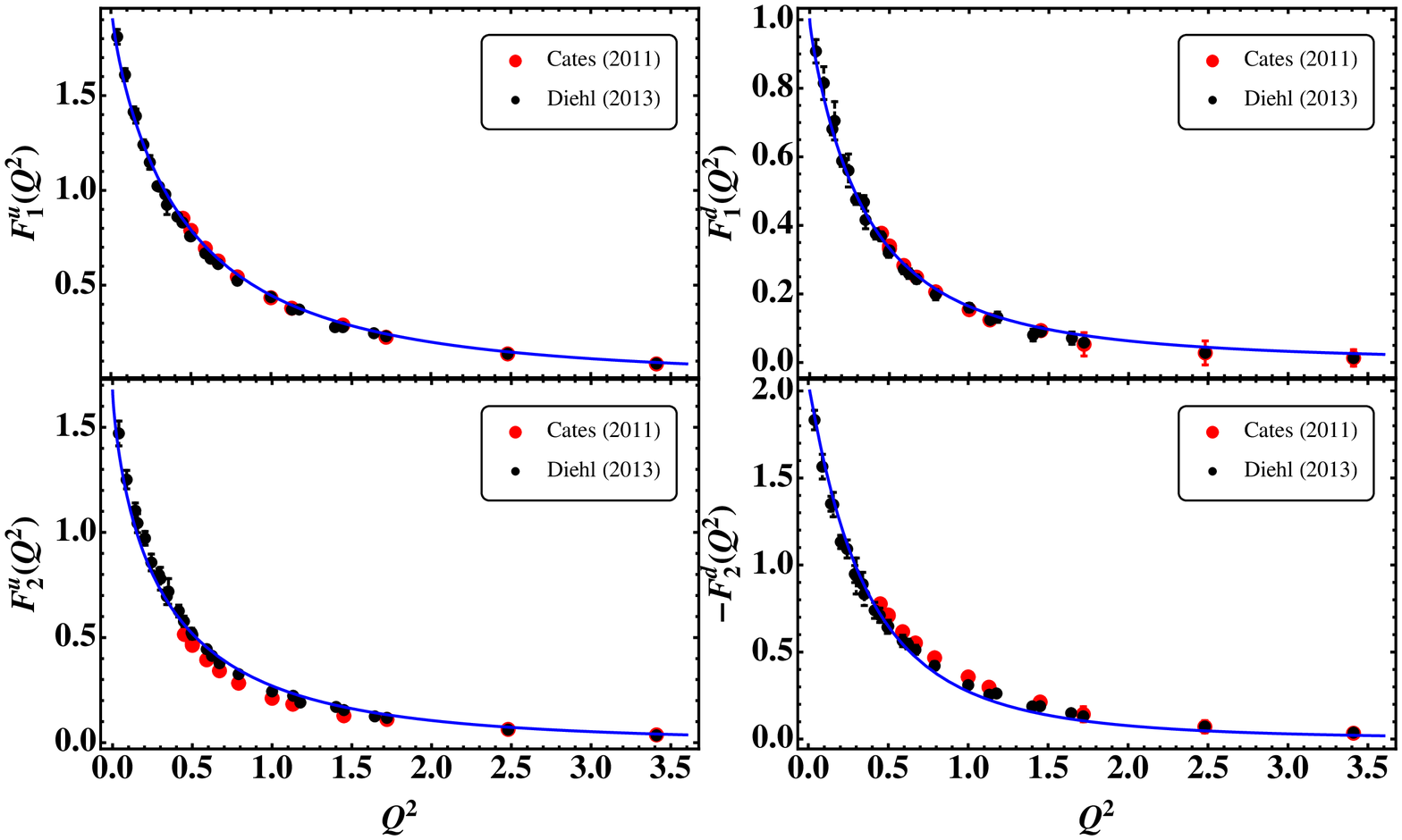}\\
\caption{ \label{q1} 
Plots of flavor form factors $F_1^u$, $F_1^d$, $F_2^u$, and $F_2^d$ with $q^2$ for up and down quarks. The experimental data points are  obtained from \cite{diehl,cates}. }
\end{center}
\end{figure}

In order to check the consistency of our approach, we have presented  the results for separate contribution of  flavor form factors for  up and down quarks in  Fig. \ref{q1}. It is straightforward to obtain the flavor decomposition of the nucleon form factors using the charge and isospin symmetry $F_j^u = 2 F_j^p +F_j^n $  and $F_j^d = F_j^p + 2 F_j^n $ $(j=1,2)$, with the normalizations $F_1^u(0) = 2$, $F_2^u(0) = \kappa_u =1.673$ and $F_1^d(0)=1,F_2^d(0)=\kappa_d = -2.033$. We have also compared  our predictions with the experimental data for flavor form factors \cite{cates,diehl}.  Cates {\it  et al.} \cite{cates}  have used  measurements on  $G^p_E$, $G^p_M$, $G^n_E$, and $G^n_M$ with the Kelly's fit  and  Diehl and Kroll \cite{diehl} have interpolated  the available data on  Sach form factors to determine the flavor form factors.  We observe that our parametrization based results are  in excellent agreement with both the experimental data sets for the individual values of flavor form factors $F_1^u$, $F_1^d$, $F_2^u$, and $F_2^d$ at the low and intermediate energy range.

\begin{figure}[htbt]
\begin{center}
\includegraphics[width=15cm,height=11cm,clip]{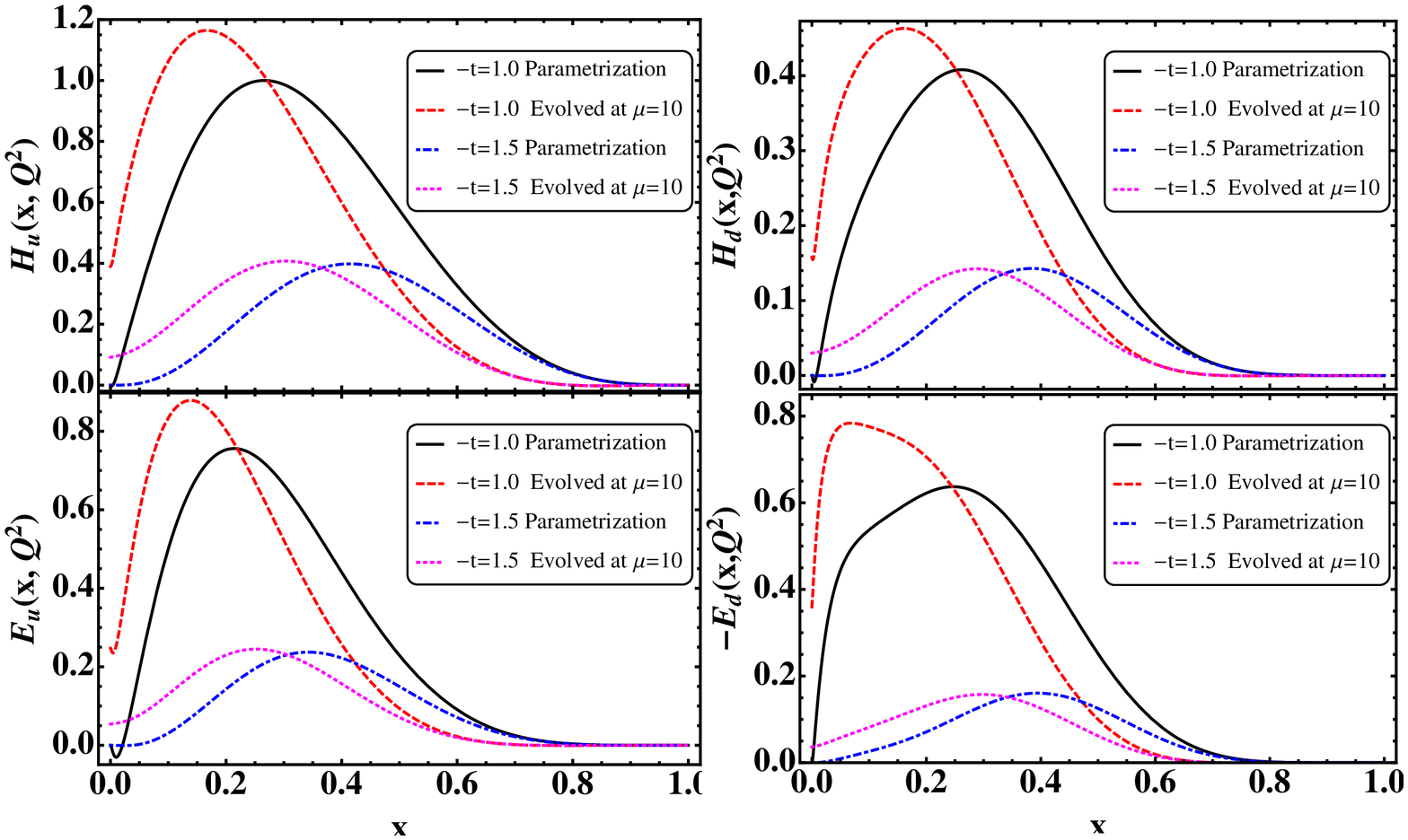}
\caption{  \label{hu1} Plots of the generalized parton distributions $H_v(x,t)$ and $E_v(x,t)$ vs $x$ for fixed values of $ -t =1, 1.5$ GeV$^2$ for up and down quark. }
\end{center}\end{figure}

For the sake of completeness, we will also discuss the  behavior of  GPDs $H_v(x, q^2)$  and $E_v(x, q^2)$  with $x$ for fixed values  $-t = 1.0, 1.5$ GeV$^2$  for the valence quarks in Fig. \ref{hu1}.  
We calculate the  up and down quark GPDs  from our parametrization form of proton  in Eq. (\ref{hp}) and neutron with charge  and sospin symmetry: $H_v^u(x, q^2)=  2 H^p(x, q^2)  +H^n(x, q^2) $ and $H_v^d(x, q^2)=  H^p(x, q^2)  + 2 H^n(x, q^2) $. From Fig. \ref{hu1}, it is clear that the qualitative behaviour of  GPDs for $x$ is same for for both up and down quarks over a wide range. In all the cases, the behavior of GPDs $H_v(x, q^2)$ and $E_v(x, q^2)$ increases with $x$ to a maximum value  and then decreases to zero.  For all cases the peak of GPDs shift towards a higher value of $x$ as the value of $q^2$ increases.

It is important to predict the scale evolution of these GPDs as various experiments give data at different energy scales \cite{dglap1,dglap2,dglap4}. The independence of physical observables from the physical scale, give the following type of DGLAP like equation for valence quark GPD $H^q (x, q^2)$:
\be
\mu^2 { {\rm d} \over  {\rm d} \mu^2} H^q (x, q^2, \mu^2)  =  \left ({\alpha_s  \over 2 \pi}  \right)\int_x^1 { {\rm d} z \over z }
 \left[ P \left( {x \over z} \right) \right]_+ H^q (z, q^2, \mu^2)\,, 
 \label{dg}
 \ee
 where $[....]_+$ is the usual ``plus regularization'' scheme for the  DGLAP evolution kernel  \cite{melin}.  The leading order quark-quark splitting function $P(z) = C_f ({1 + z^2 \over 1-z}) $ gives the probability of a quark after being  radiating a gluon is left with momentum fraction $z$  of the original momentum. We have solved the Eq. (\ref{dg})  using numerical methods \cite{Sharma2016} and absorbed the uncalculable perturbative effects into the  evolved GPDs.  In order to understand the implication of the  DGLAP evolution on GPDs, we have also presented the results with scale parameter $\mu =10$ GeV.  For the evolved GPDs, peak shift towards a lower value of $x$ for the higher values of scale parameter $\mu$ as the probability of a gluon being radiated is higher at large values of $x$, hence the distribution shift towards lower value of $x$. 
 

\section{Transverse charge and magnetization densities}
\label{transcha}

In the previous section, we have established the $q^2$-dependence of GPDs and successfully calculated the Sach form factors and  the individual contribution of  flavor FFs and GPDs for the up and down quarks.  In this section, we will calculate the nucleon transverse densities in  the parametrization approach. The transverse charge and magnetization densities for the unpolarized nucleon are defined as the two-dimensional  Fourier transform of the electromagnetic form factors $F_{1}$ and $F_{2}$  with respect to  transverse momentum transfer ($q_\perp$). These densities are  further related to the first order moment of GPDs in the transverse impact parameter space.  These density functions are interpreted as the probability density for finding a quark with a longitudinal momentum fraction ($x$) located at the transverse position ($b_\perp$)  inside the nucleon.

\subsection{ Transverse charge  densities inside the unpolarized nucleon}

The  charge density $(\rho_{ch}^N)$  in transverse impact parameter space for the unpolarized nucleon is expressed as
\be \rho_{ch}^N (b_\perp) = \int { {\mathrm d^2} q_\perp \,  \over (2 \pi)^2 }  F_1^N(q^2) e^{\iota q_\perp \cdot b_\perp}  =  \int_0^1 {\mathrm d} x\, q^N(x, b_{\perp}) \,,  \label{ch} \ee
where $q^N(x, b_\perp)$  is impact parameter GPD  given by the Fourier transform of momentum space GPD $H^N(x,q^2)$ with respect to the  momentum transfer ($ q_{\perp}$).  For zero skewness, the momentum transfer is only in the transverse direction, thus $q^N(x,b_\perp)$ gives the transverse distribution of the partons in impact parameter space:
\bea q^N(x,\, b_\perp)  &=& {1 \over {(2 \pi) }^2} \int {\mathrm d}^2 q_{\perp} e^{\iota  q_{\perp} \cdot b_{\perp } } H^N(x, \, q^2) \,.  \eea

One can similarly calculate the magnetization density $(\rho_m^N)$  in transverse impact parameter space defined as the Fourier transform of the Pauli form factors
 \bea {\tilde \rho_{m}^N (b_\perp)} = \int { {\mathrm d^2} q_\perp \,  \over (2 \pi)^2 }  F_2^N(q^2) e^{\iota q_\perp \cdot b_\perp} =  \int_0^1 {\mathrm d} x\, {\cal E}^N(x, b_{\perp})\,, \eea
where  \bea {\cal E}^N(x,\, b_\perp) &=& {1 \over {(2 \pi) }^2} \int {\mathrm d}^2  q_{\perp} e^{- b_{\perp}. q_{\perp } } E^N(x, \, q^2) \,,\eea however, the true  anomalous magnetization density in the transverse plane is expressed as \bea  \rho_m^N(b_\perp) &=& -b{ \partial  {\tilde \rho_m^N (b)} \over  \partial b} \,, \nonumber \\ &=& b \int_0^{\infty} \frac{{\mathrm d} q_\perp} {2 \pi} q^2 J_1(q b)   \int_0^1 {\mathrm d} x \,  {\cal E}^N (x\,, b_\perp)\,.\eea

\begin{figure}[htbt]
\begin{center}
\includegraphics[width=16cm,height=15cm,clip]{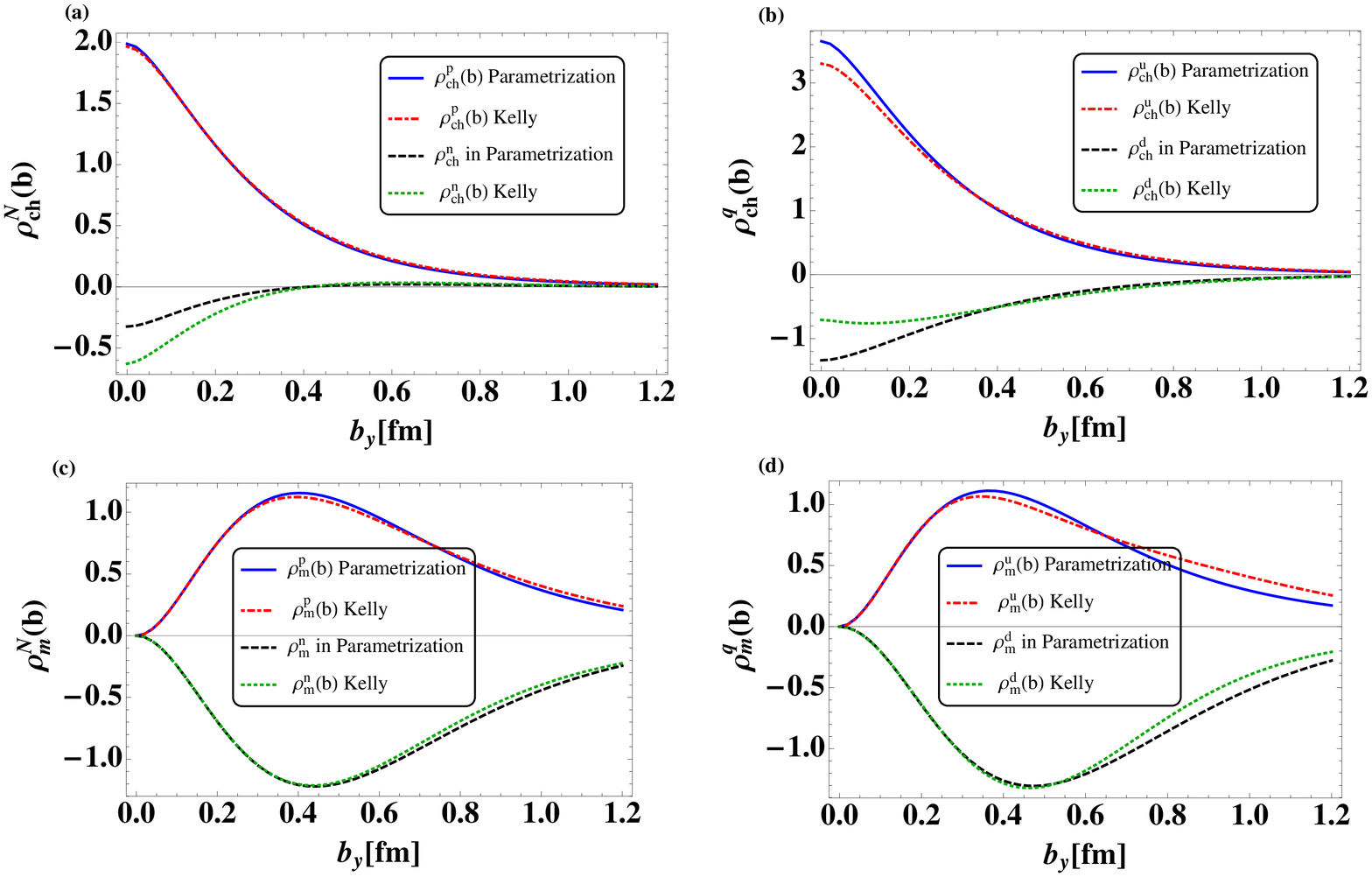}
\caption{Plots of transverse charge $\rho_{ch}^N (b)$  with $b_y$ for (a) proton and neutron (b) up and down quark. Plots of transverse magnetization density $\rho_{m}^N (b)$  with $b_y$ for (c)  proton and neutron (d) up and down quark.
\label{mp1}}
\end{center}
\end{figure}

In order to obtain information about the transverse charge and magnetization densities inside the unpolarized nucleon, we plotted them in Fig. {\ref{mp1}}.  In Fig. \ref{mp1} (a), we have plotted charge density  for proton and Neutron. These plots give the complete spatial information about the nucleon. A cursory look at the  plots reveal that charge distributions is maximum at the center and then decreases with increasing value of the impact parameter $b$. In particular, the proton charge density  for unpolarized nucleon has a large positive value at the center of the core which decreases further with the increase in $b_\perp$. Plots for the neutron charge density support the large negative value at the center of core and then a positive contribution at the intermediate $b$ values  and then decreases. 

There is no experimental data directly available for the charge and magnetization densities, we compare our parametrization approach results with Kelly model \cite{kelly}. Kelly has proposed a parametrization for the nucleon Sachs form factors using Laguerre-Gaussian expansion. The overall behaviour for the charge densities are same as the  Kelly's model, however the matching of charge densities is perfect for the case of proton but differ slightly for the case of neutron. A phenomenological analysis of existing data to determine the charge density of the proton and neutron has been discussed in Ref. \cite{miller2007,*Miller:2010nz}  in a model independent way. The predictions of Ref.  \cite{miller2007,*Miller:2010nz} also highlight the negative value of 
 neutron parton charge density at  center, so that the square of the transverse charge radius is positive, same as predicted in parametrization approach.

The decompositions of the transverse charge and magnetization densities for the nucleon can be defined in a similar way as the EFFs. In terms of two flavors one can write down the transverse charge and magnatization densities for the up and down quark as $\rho_{ch,m}^u = 2 \rho_{ch,m}^p +\rho_{ch,m}^n $  and $\rho_{ch,m}^d = \rho_{ch,m}^p + 2 \rho_{ch,m}^n $.
In Fig. \ref{mp1}(b), we have presented the result for charge density distribution for up and down quark, respectively.  The transverse charge density is positive for the up quark and negative for down quark.  The magnitude of charge density for  up quarks is found to be larger the down quark. Authors in the Ref. \cite{miller2007,*Miller:2010nz}  also observe a large value of the up quark density as compared to the down quark density.


\subsection{ Transverse charge  densities inside the polarized nucleon}

Recently, Carlson and Vanderhaeghen have further generalized the charge density  for transversely polarized nucleon in  light-front form \cite{carl}. So it is interesting to investigate the charge density in transverse plane  for  transversely polarized nucleon expressed as:
\bea \rho_T^N (b_{\perp }) &=&  \rho_{ch}^N(b_{\perp }) - {\sin (\phi_b - \phi_s)   \over 2 M_N b}  \rho_{m}^N (b_{\perp })  \,. \label{rhoT}  \eea
We consider a nucleon polarized in the $xy$ direction with the transverse polarization $S_\perp = \cos \phi_s {\hat x} + \sin \phi_s {\hat y}$ and  transverse impact parameter $b_\perp =  b( \cos \phi_b {\hat x} + \sin \phi _b {\hat y})$.  The first term in Eq.(\ref{rhoT}) is the unpolarized charge density and   second term gives the deviation from circular symmetry for unpolarized charge density and  depends on the orientation of $b_\perp$ relative to the transverse spin direction $S_\perp$.

\begin{figure}[htbt]
\begin{center}
\includegraphics[width=16cm,height=15cm,clip]{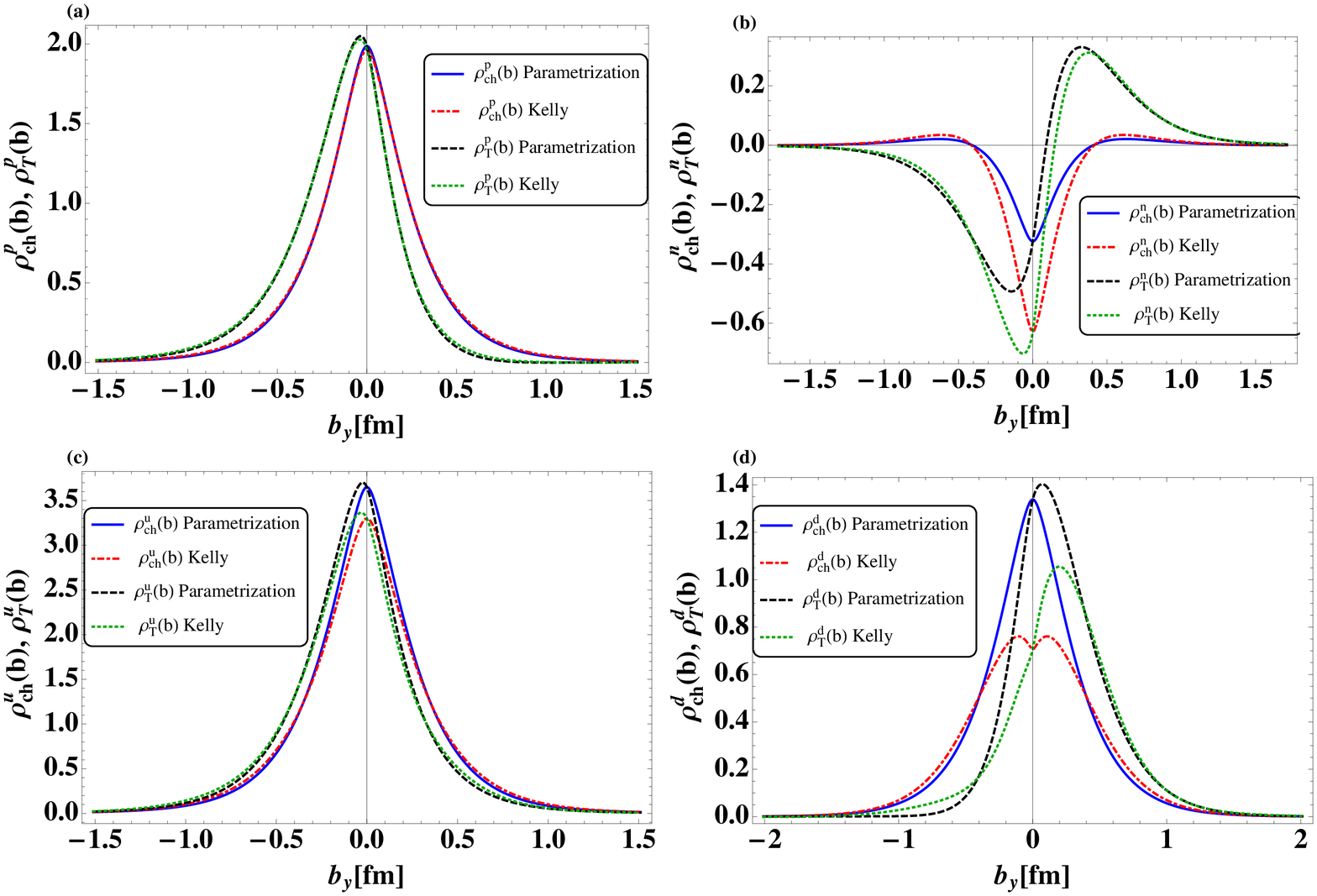}
\caption{Plots of transverse charge densities  $\rho_{ch}^N (b)$ and  $\rho_{T}^N (b)$  with transverse impact parameter space $b_y$ for  the unpolarized and transversely polarized (a) proton  (b) neutron (c)  up quark, and (d) down quark. The transverse polarization is taken along the $x$ direction.
\label{cmp}}
\end{center}
\end{figure}

We have investigated the  transverse charge densities for transversely polarized proton and neutron in the parametrization approach and presented the results in Fig. \ref{cmp}(a)-(b).  We observe that the polarized proton charge density has a large positive value at the center of the core which decreases further. Polarized neutron charge density reveal a large negative value at the center of core and then a positive contribution at the intermediate $b$ values  and then decreases. This trend in the behaviour of charge densities is same as for the unpolarized nucleon densities. 
Apart from this, it is also important to mention that the transverse charge densities for the proton polarized transversely along the positive $x$ direction ($\phi_s=0$) are distorted in the negative $b_y$ direction.  In  the case of neutron transversely polarized along the $x$ axis, the negative charge is shifted to the negative $b_y$ direction and the positive charge move towards to the positive $b_y$ direction. We have also compared our $q^2$-dependent results with the Kelly approach. The parametrization results for the polarized transverse charge densities  are in agreement with the Kelly predictions for the case of proton, however the results differ slightly from the Kelly distribution  for the neutron.

In  Fig. \ref{cmp}(c)-(d), we have also investigated the up and down  quark transverse charge densities inside the unpolarized and transversely polarized nucleon in the parametrization method. The transverse charge densities ($\rho_{ch}^{u/d} (b_\perp)$ and $\rho_{T}^{u/d} (b_\perp)$) for up and down quark  can be obtained using the flavor form factors.  A cursory look at the graphs reveal that the up quark transverse charge density inside the transversely polarized nucleon is shown to be shifted to the negative $b_y$ direction, while  it is distorted in the positive $b_y$ direction for the down quark. Parametrization  results for the up quark unpolarized and polarized transverse charge densities  matches significantly with  Kelly approach, however the deviation is more prominent in the case of down quark. Our  predictions  are also  in significant agreement with the other phenomenological models in the literature \cite{Burkardt:2000za,*Burkardt:2002hr,csqm,lattchargeden,carl, neetika}. Therefore,  parametrization approach with the $q^2$-dependence and perturbative evolution effects play an important role in understanding  the hadron structure.


\section{Summary and conclusion}
\label{conc}
In this work, we have introduced a phenomenological parametrization  for $q^2$-dependence of the  proton and  neutron GPDs  based on Gaussian ansatz  along with the `MSTW2009'  results for the quark distribution function \cite{Martin:2009iq}. A simple form of parametrization with minimum number of free parameters gives a sufficiently good description of the experimental data on Sach form factors and the ratio of electric to magnetic form factors over a  wide range of  momentum transfer. We have also investigated the flavor decompositions of the nucleon form factors for up and down quarks using the charge and isospin symmetry.  The  parametrization method results for flavor form factor are in agreement with the experimental data.  Since the first moments of the GPDs give the EFFs, we have investigated the GPDs $H$ and $E$ for up and down quarks in nucleon  and ensured their stability under the DGLAP evolution.

Further, We have  presented a detail comparison of transverse  charge densities for both the unpolarized and transversely polarized nucleon and their  flavor decompositions into up and own quark.  Though we have considered only valence quarks contribution in the paramerization approach, it is interesting to note that the qualitative  behaviour of transverse charge densities  is same as the Kelly's approach. The unpolarized densities are axially symmetric in the transverse plane, while densities become distorted for the transversely polarized nucleon. If the nucleon is polarized along positive $x$ direction, the densities get shifted towards negative $y$ direction. The transverse distortion as well as the deviation from experimental results are more prominent in the case of down quark than up quark. We showed that combining the  parametrization method with the  $q^2$-dependence  give the effective model for understanding the three-dimensional structure of the nucleon in the nonperturbative region.

\section*{Acknowledgements}
 This work is supported by Department of Science and Technology, Government of India (Grant No. SR/FTP/PS-057/2012).

 \bibliographystyle{apsrev4-1}
 
\bibliography{bib}

\end{document}